\documentclass{birkart04}
 \theoremstyle{definition}
 
 \theoremstyle{remark}

 \numberwithin{equation}{section}

 \DeclareMathSymbol{\R}{\mathbin}{AMSb}{"52}

\begin{document}
%--------------------------------------------------------------------------
% editorial commands: to be inserted by the editorial office
%
%\firstpage{1}
%\volume{228}
%\DOI{003-0001}
%\seriesextraline{This is the Concrete Title of this Book\br H.E. Rowe and S.T.C. Wore, Eds.}
%\commby{inhouse GRRR}
\title[Continuum models for surface growth]
 {Continuum models for surface growth}
%----------Author 1
\author[Martin Rost]{Martin Rost}

\address{%
Bereich Theoretische Biologie\\
Insitut f\"ur Zellul\"are und Molekulare Botanik\\
Kirschallee 1, Universit\"at Bonn\\
53115 Bonn, Germany}

\email{martin.rost@uni-bonn.de}

%\thanks{}
%----------Author 2
%\author{A Second Author}
%\address{The address of\br
%the second author\br
%sitting somewhere\br
%in the world}
%\email{dont@know.who.knows}


\date{March 26, 2004}
%\dedicatory{To my parents}
%%% ----------------------------------------------------------------------

\begin{abstract}
As an introductory lecture to the workshop an overview is given over
continuum models for homoepitaxial surface growth using partial
differential equations (PDEs). Their {\em heuristic derivation} makes
use of inherent symmetries in the physical process (mass conservation,
crystal symmetry, \dots) which determines their {\em structure}. Two
examples of applications are given, one for large scale properties,
one including crystal lattice discreteness. These are: (i) a
simplified model for {\em mound coarsening} and (ii) for the
transition from {\em layer-by-layer} to {\em rough growth}. Virtues
and shortcomings of this approach is discussed in a concluding
section.
\end{abstract}

%%% ----------------------------------------------------------------------
\maketitle
%%% ----------------------------------------------------------------------
%\tableofcontents

\section{Introduction}
Crystal growth by Molecular Beam Epitaxy (MBE) involves processes on
quite different physical scales, from atomic to micrometer scales. To
capture the large scale features it may be appropriate to represent
the surface as a continuous height field $h({\bf x},t)$ above a set of
so called base points ${\bf x}$ where $h : ({\bf x},t) \in \R^2 \times
\R \to \R$. The surface dynamics is then modelled by a parabolic PDE
for $h({\bf x},t)$, which often is a starting point interesting
mathematical problems, and in some cases for rigorous results
\cite{KO02,KY03}, which were an important part of this workshop.

Their mathematical tractability somehow contrasts with the lack of
rigour in their derivations, as we shall see below. It would be nice
to ``prove'' a certain PDE to be the appropriate large scale model for
some underlying more elementary model, typically a stochastic model of
the solid--on--solid type. This would connect the result for large
scales with the microscopic model which is much closer to physical
reality.

Nevertheless, often this chain can be closed in a heuristic sense, and
this lecture presents some examples. It is organised as follows:
First, the heuristic procedure of deriving a continuum model is
motivated and its structure is explained. Then, two examples of
applications are presented, covering in some sense extreme cases of
applicability of continuum models; (i) coarsening of mound patterns in 
homoepitaxy and (ii) damping of width oscillations in the transition
from layer--by--layer to rough growth. Last, the general applicability
of continuum models to MBE growth is discussed, in particular those
features of discreteness of the underlying process which can only be
covered incompletely by a continuum model. Open questions in the
derivation which have not been addressed so far, but appear to be
solvable, are mentioned here.

The main purpose of this introductory lecture is to recollect the
physical motivation of the class of models which deal with the largest
spatial and temporal scales involved in MBE growth, and to give some
intuition to other more mathematical approaches.

\section{Derivation and structure}

\subsection{General background}
\begin{figure}[!h]
\begin{center}
\includegraphics[width=8cm]{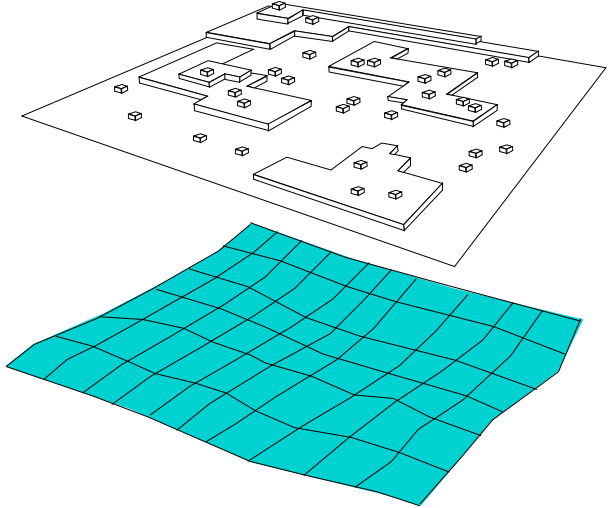}
\end{center}
\caption{Sketch of the hight field representation for surface
  configurations. Upper: one surface configuration out of an ensemble
  in local steady states. Lower: a smooth height field $h({\bf x},t)$
  as a representation.}
\label{Fig:Ensemble}
\end{figure}

Continuum models for surface growth are comparable to the
hydrodynamic limit of microscopic models of fluid dynamics. Behind the
continuous macroscopic height field $h({\bf x},t)$ one has to imagine
an ensemble of microscopic configurations which are locally in a
steady state. Figure~\ref{Fig:Ensemble} illustrates this
relationship. The local state is subject to changes on time scales,
ideally longer than the relaxation times into those states, which
leads to the dynamics of the height field.

The structure of the PDE governing the dynamics of $h({\bf x},t)$
reflects different elementary processes and their symmetries. In
growth from molecular beam epitaxy there is
\begin{itemize}
\item a deposition flux $F_{\{ h \}}({\bf x},t)$, i.e., the deposited
  volume per time and surface area,
\item possibly an evaporation loss $E_{\{ h \}}({\bf x},t)$, again
  volume per time and area,
\item a surface current ${\bf J}_{\{ h \}}({\bf x},t)$ on terraces
  and along steps, transported volume per time and length, i.e., cross
  section on the surface.
\end{itemize}
Multiplied with a volume density $\rho$ these quantities become mass
fluxes and currents. All three depend on the actual surface
configuration $h$, e.g., on local slopes, their orientations and
curvatures. Elastic interactions and terrace adatom detachment with
subsequent redeposition couple the dynamics at a surface point {\em
  nonlocally} to surface configurations at other places. There may
also be an {\em explicit} space dependence beyond the influence of the
height configuration, e.g., for a non--uniform deposition with pulsed
flux.

Deposition, evaporation, and surface current are combined into a {\em
  continuity equation}
\begin{equation}
\frac{\partial}{\partial t} \; h({\bf x},t) = F_{\{ h \}}({\bf x},t) -
\nabla \cdot {\bf J}_{\{ h \}}({\bf x},t) - E_{\{ h \}}({\bf x},t).
\label{eq:surfdyn}
\end{equation}
Note that small local slopes are assumed, such that any geometric
prefactors to the divergence of the surface current typically are
neglected.

A full derivation from local steady states in the sense postulated
above is of course impossible in almost any case, apart from some
one-dimensional examples \cite{ASEP}. One therefore has to rely on
heuristically justified ({\em reasonable}) assumptions for the dynamic
quantities involved. The connection to the more detailed models on
smaller scales is given by length scales which influence the surface
morphology and become visible. Explained more precisely in the other
introductory lectures of this conference \cite{BiehlOW,KrugOW} they
are
\begin{itemize}
\item $a$, the lattice spacing of the crystal.
\item $\ell_{\rm D}$, the diffusion length, i.e., the ``typical''
  distance of nucleation sites on a terrace, with a non--trivial
  dependence on the deposition rate $F$ and adatom diffusivity $D$.
\item$\ell_{\rm ES}$, the Ehrlich--Schwoebel length, i.e., a measure
  for the reflectivity of the barrier at a downward step. Roughly
  speaking, an adatom located right at a downward step is equally
  likely to hop downward as it is to reach the next upward step, if
  it lies at a distance $\ell_{\rm ES}$.
\item$\ell_{\rm step}$, the average distance between two truly
  ascending at a vicinal surface of slope $a/\ell_{\rm step}$ without
  counting islands between them.
\end{itemize}

\subsection{Surface current}
{\bf Ehrlich--Schwoebel effect on terraces}\\
As was realised quite early the most important ingredient of
Eq.~(\ref{eq:surfdyn}) is a non--equilibrium current on surface
terraces induced by Ehrlich--Schwoebel (ES) barriers
\cite{Ehrlich66,Schwoebel66,Siegert97,Villain91}.

For small average slopes ${\bf m} = \nabla h$ with $|{\bf m}| \ll
a/\ell_{\rm D}$ or $\ell_{\rm D} \ll \ell_{\rm step}$, the ES barrier
induces a net uphill current ${\bf J}_{\rm EST} \sim F {\bf
  m}$. Inserted into the surface current in Eq.~(\ref{eq:surfdyn})
this results in a {\em destabilising} term for the height equation
$\partial_t h = - c \nabla \cdot \nabla h + \dots$.

Of course one would like to know the functional dependence of the
surface current ${\bf J}_{\rm EST}$ beyond this approximation for
small slopes. Following an argument of Siegert for $1 \! + \! 1$
dimensions \cite{Siegert97} there must also be surface orientations
with stabilising surface current: Think of ${\bf
  J}_{\rm EST}$ as a vector field on the tangent bundle of the unit
sphere of possible surface orientations. By the ES instability all
high symmetry orientations have ${\bf J}_{\rm EST}$ pointing outward
around a zero. By continuity of ${\bf J}_{\rm EST}$ as a function of
the orientation there must be zeroes with the current field pointing
inward to the zero, i.e., orientations stabilised by the surface
current.

These qualitatively fundamental properties, the existence of unstable
and stable surface orientations, can be cast into a heuristically
postulated function ${\bf J}_{\rm EST} ({\bf m})$.
\begin{figure}[!ht]
\begin{center}
\includegraphics[width=5cm]{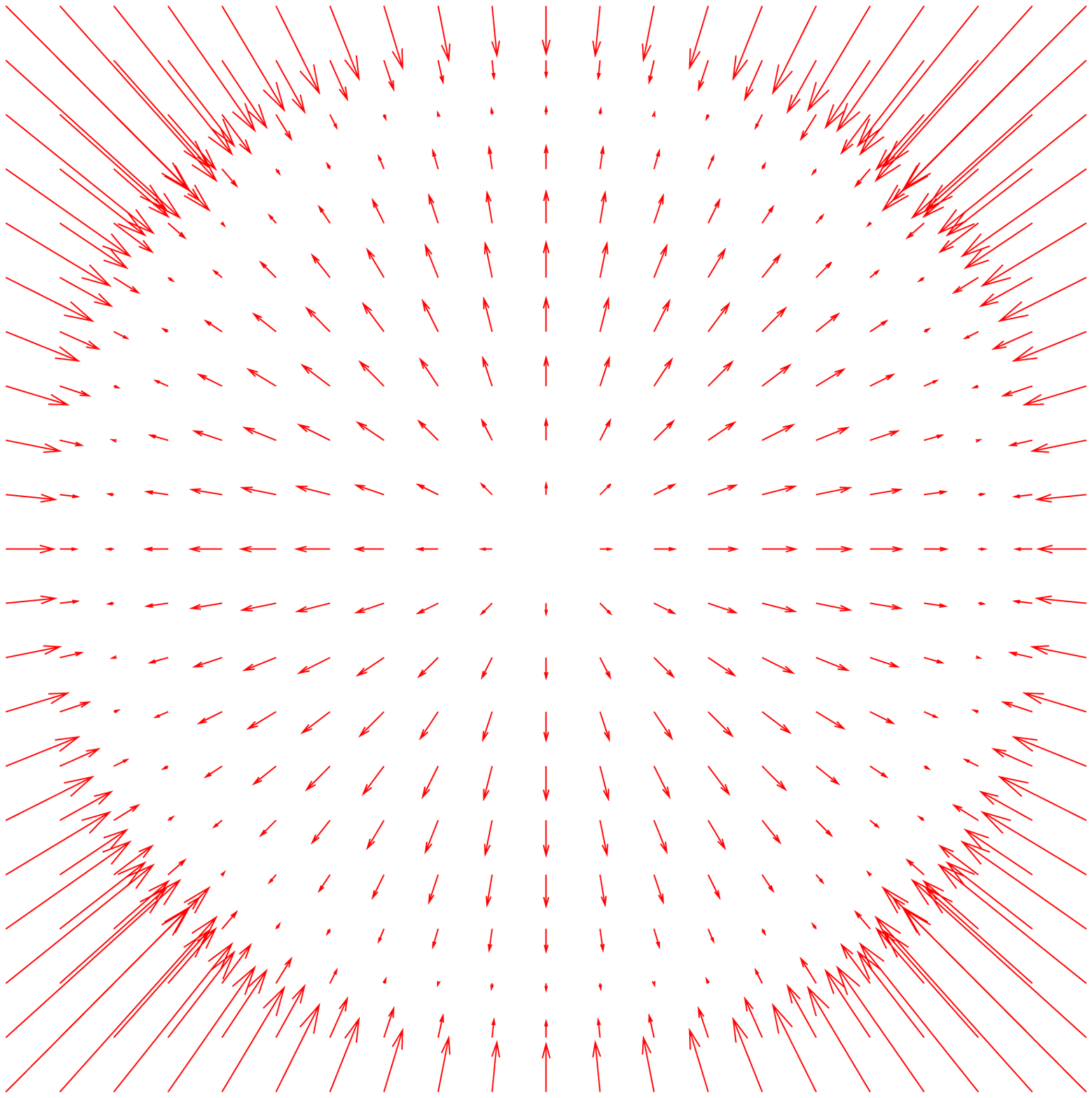}
\includegraphics[width=5cm]{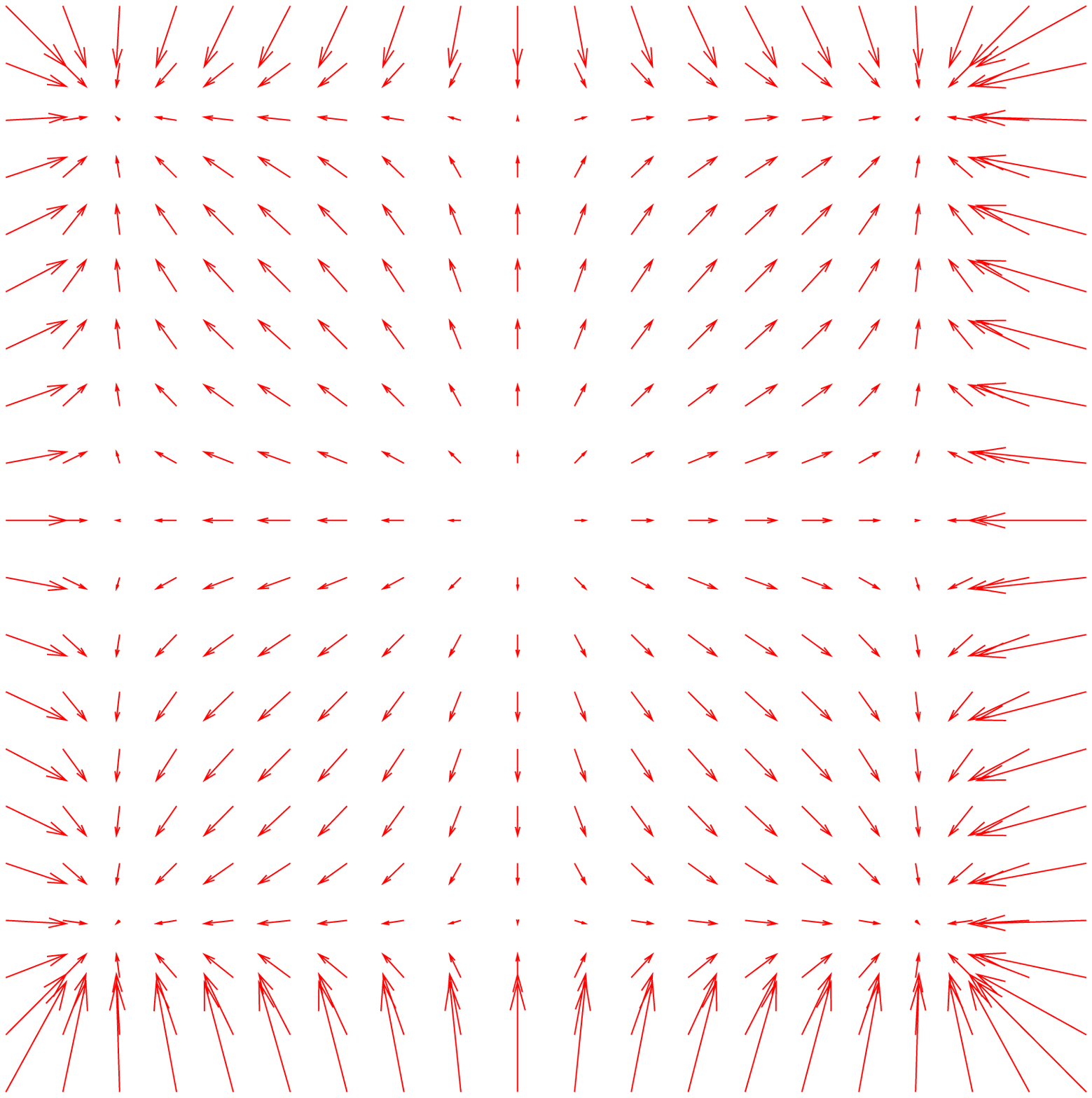}
\end{center}
\caption{Examples of heuristic choices for nonequilibrium terrace
  Ehrlich-Schwoebel surface current ${\bf J}_{\rm EST} ({\bf m})$.
  Left: in--plane isotropy, ${\bf J}_{\rm EST} ({\bf m}) = (1 - |{\bf
  m}|^2) \; {\bf m}$, continuous set of stable slopes $\{ {\bf m}\; |
  \; |{\bf m}| \! = \! 1 \}$. Right: anisotropy, four--fold symmetry,
  $J_{{\rm EST},i} ({\bf m}) = (1 - m_i^2) \; m_i$ with $i = x,y$,
  discrete set of stable slopes at $( \pm 1, \pm 1)$.} 
\label{Fig:ES}
\end{figure}
Figure \ref{Fig:ES} shows two such examples, one isotropic within the
crystal terrace --- a somewhat artificial but nevertheless
theoretically interesting case, one isotropic, with four--fold ($\pi
/2$) rotational symmetry.
\medskip

\begin{figure}[!ht]
\begin{center}
\includegraphics[width=8cm]{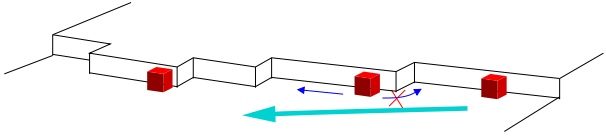}
\end{center}
\caption{The ES effect at step kinks favours adatom incorporation from
  one side and induces a net current along the step.} 
\label{Fig:ESS}
\end{figure}

\noindent
{\bf Ehrlich--Schwoebel effect at steps}\\
The Ehrlich--Schwoebel effect also has an influence on adatom
diffusion along steps, allowing for attachment of atoms to kinks
preferentially from one side, as is sketched in Fig.~\ref{Fig:ESS}.
It is one important mechanism causing a step meandering instability on
growing vicinal surfaces \cite{Orsogna99,Kallunki02}.
Politi and Krug derived the contribution ${\bf J}_{\rm ESS}$ of
non--equilibrium current along steps to the total surface current for
a surface with given average slope which is assumed to be constant on
a sufficiently large region such that steps have a well defined
average density and orientation with respect to the underlying crystal
lattice \cite{PK00}.
\medskip

\noindent
{\bf Curvature relaxation}\\
Clearly on small scales the destabilising ES effects encountered in
this section so far are balanced. Too thin protrusions are forbidden
by their cost in surface free energy. For thermal relaxation towards
equilibrium the corresponding continuum dynamics are straightforward
to derive. The surface free energy
\begin{equation}
\mathcal{F} = \int \gamma(\nabla h) \; \sqrt{1 + (\nabla h)^2} \; d^d x
\end{equation}
weights each surface element by its orientation dependent
contribution $\gamma(\nabla h)$. The gradient in chemical potential
$\mu = \delta\mathcal{F}/\delta h$ drives the relaxational surface
current which is also proportional to the (tensorial and orientation
dependent) adatom mobility {\boldmath$\sigma$}$(\nabla h)$. A linear
expansion around a surface of constant slope $h = {\bf m \cdot x}$
small enough to again neglect geometric effects
\cite{Mullins59,Villain91} yields
\begin{equation}
{\bf J}_{\rm CURV} = \mbox{\boldmath$\sigma$} \! \cdot \! \nabla \;
  \nabla  \! \cdot  \! \left[ \gamma \; {\bf 1} + (\nabla_{\! \bf m}
  \! \nabla_{\! \bf m} \gamma )\right]  \! \cdot  \! \nabla h.
\label{eq:mullins}
\end{equation}
The expression in square brackets is generally called the surface {\em
  stiffness}
\begin{equation}
\tilde{\mbox{\boldmath$\gamma$}} = \gamma \; {\bf 1} + \nabla_{\! \bf
  m} \! \nabla_{\! \bf m} \gamma
\end{equation}
where $\nabla_{\! \bf m}$ denotes derivation with respect to the slope
dependence of $\gamma$ and ${\bf 1}$ is the unity matrix.

Strictly speaking this derivation applies only to {\em equilibrium
  relaxation}. For a far from equilibrium process like MBE growth it
does not make sense to define a surface free energy per area
$\gamma(\nabla h)$, although the ``true'' contribution of the surface
current due to relaxation by curvature most likely has a functional
form of that type. The identification of these contributions and the
correct derivation of the corresponding continuum form remains an open
question.

\subsection{Deposition, desorption and noise}
For most applications it will be correct to assume a {\em deposition}
intensity $F$ constant in time and space. Interesting cases with {\em
  explicit} space and time dependence are (i) pulsed molecular beams,
used to decrease the second layer island nucleation rate, and (ii)
geometric inhomogeneities if the beam does not reach every part of the
surface under the same angle. The most important mechanism for {\em
  implicit} deposition heterogeneity is steering, where the incident
atoms are deflected by attractive interaction with surface features
such as islands and mounds, and effect which is of course most
prominent under grazing incidence \cite{Dijken99,Yu04}.

At sufficiently high temperatures {\em desorption} may become an
important effect. Terrace adatoms then will be most susceptible to
desorption at a rate $1/\tau$, during their diffusion time before they
reach the stronger binding steps. To a first approximation the
desorption rate depends on the average distance to the next reachable
step and therefore on the surface slope: the higher the slope, the
more steps are present and the lower the desorption loss. Within an
ansatz \`a la Burton--Cabrera--Frank \cite{BCF} one can relate the
desorption length $x_{\rm s} = 2 a \sqrt{D \tau}$, diffusion length
$\ell_{\rm D}$, and slope $\nabla h = a/\ell$
\cite{Smilauer99,Villain91} to an expression for the desorption loss
in volume per time and surface area
\begin{equation}
E_{ \{ h \} }({\bf x},t) = F \; \left[ (x_{\rm s}/\ell) \tanh (\ell/x_{\rm
    s}) - 1 \right] \approx F \; \frac{(\ell_{\rm D}/x_{\rm
    s})^2/3}{1+(\nabla h)^2}.
\end{equation}

There are three sources for {\em noise} in MBE growth \cite{Wolf95}:
(i) Deposition is a stochastic process of single events, so $F$ has to
be complemented by a ``shot noise'' term $\xi({\bf x},t)$ with
statistics
\begin{equation}
\langle \xi({\bf x},t) \rangle = 0 \; \; \; \; \mbox{and} \; \; \; \;
\langle \xi({\bf x},t) \xi({\bf x}',t') \rangle = a^3 F \; \delta({\bf
  x} \! - \! {\bf x}') \; \delta(t \! - \! t').
\end{equation}
(ii) Adatom diffusion also consists of single events, and to a first
approximation surface currents are blurred by a ``diffusion noise''
term {\boldmath$\eta$}$({\bf x},t)$ with statistics
\begin{equation}
\langle \mbox{\boldmath$\eta$}({\bf x},t) \rangle = 0 \; \; \; \;
\mbox{and} \; \; \; \; \langle \mbox{\boldmath$\eta$}({\bf x},t) \cdot
\mbox{\boldmath$\eta$}({\bf x}',t') \rangle = \ell_{\rm D} F \;
\nabla^2 \delta({\bf x} \! - \! {\bf x}') \; \delta(t \! - \! t').
\end{equation}
(iii) Nucleations also occur at random but the precise form of this
contribution remains an open question. There are nontrivial
correlations in time, as new islands tend to nucleate close to centres
of previously nucleated islands. The space dependence of the
nucleation noise correlator $\langle \nu({\bf x},t) \nu({\bf x}',t')
\rangle$ is expected to have the form $(\nabla^2)^2 \delta({\bf x} \!
- \! {\bf x}')$ \cite{Somfai96}.

\subsection{Structure and symmetry of continuum equations for MBE}
Heuristic surface currents as in the example in Fig.~\ref{Fig:ES} lead
to a symmetry in the resulting dynamical equation which can be used
for evaluation, but is too strong a restriction compared to the ``real''
dynamics and may lead to erroneous results.

If ${\bf J}_{\rm EST}$ uniquely depends on the slope ${\bf m}$ and not
on derivative of its functions then crystal symmetry imposes
${\bf J}_{\rm EST}({\bf m}) = - {\bf J}_{\rm EST}(-{\bf m})$. The
resulting term of surface dynamics is invariant under the transformation
$h \to -h$ and so will be the surface configurations obtained from
that. Breaking of this symmetry, e.g., by terms as $\nabla f\left( (
\nabla h)^2 \right)$ in ${\bf J}$ or an evaporation term $e(\left(
(\nabla h)^2 \right)$ do in fact change the characteristics of the
dynamics substantially \cite{Smilauer99,MG00}

One can take advantage of this simple slope dependence if there is a
scalar function $V({\bf m})$ such that ${\bf J}_{\rm EST} = -
\nabla_{\bf m} V({\bf m})$. Stable zeroes of ${\bf J}_{\rm EST}$ then
are minima of $V({\bf m})$. If additionally the curvature relaxation
is of a simple ``scalar'' form {\boldmath$\sigma \cdot \tilde \gamma$}
$=$ $K {\bf   1}$ there is a Lyapunov functional of the dynamics
\begin{equation}
\mathcal{F}  \{ {\bf m} \} = \int \left[ \frac{K}{2} (\nabla {\bf m})^2 +
  V({\bf m}) \right] \; d^2x
\end{equation}
and the surface dynamics Eq.~(\ref{eq:surfdyn}) can be written for the
field of slopes ${\bf m} = \nabla h$ as
\begin{equation}
\partial_t {\bf m} = \partial_t (\nabla h) = \nabla \; \nabla \cdot
\frac{\delta \mathcal{F}}{\delta {\bf m}}.
\label{eq:modelb}
\end{equation}
If ${\bf m}$ is interpreted as a two--dimensional magnetic order
parameter this from compares to conserved (model B) relaxational
dynamics, with the only restriction that it remains a conservative
field at all times, $\nabla \wedge {\bf m} = 0$. Facets in the surface
are regions of constant slope, and appear as domains of constant
``magnetisation'', edges between them are domain walls. As compared to
its magnetic analogon, here the parallel component of ${\bf m}$
remains constant across a domain wall, which in the case of discrete
stable zeroes of the current function (right panel of
Fig.~\ref{Fig:ES}) restricts the possible domain wall orientations and
imposing constraints on the relaxational dynamics
\cite{MG00,Siegert97,S98}.

\begin{figure}[!ht]
\begin{center}
\includegraphics[width=5cm]{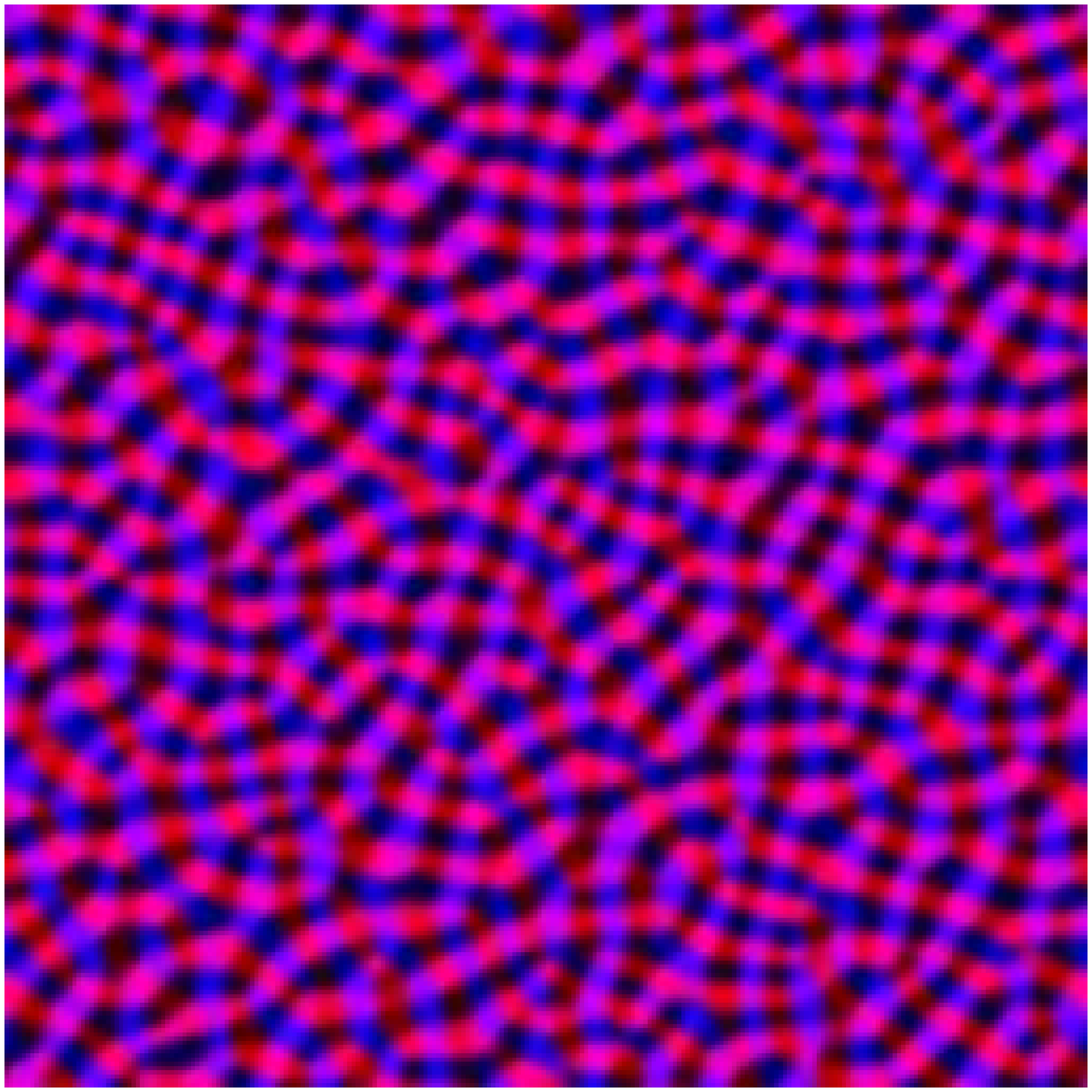}
\includegraphics[width=5cm]{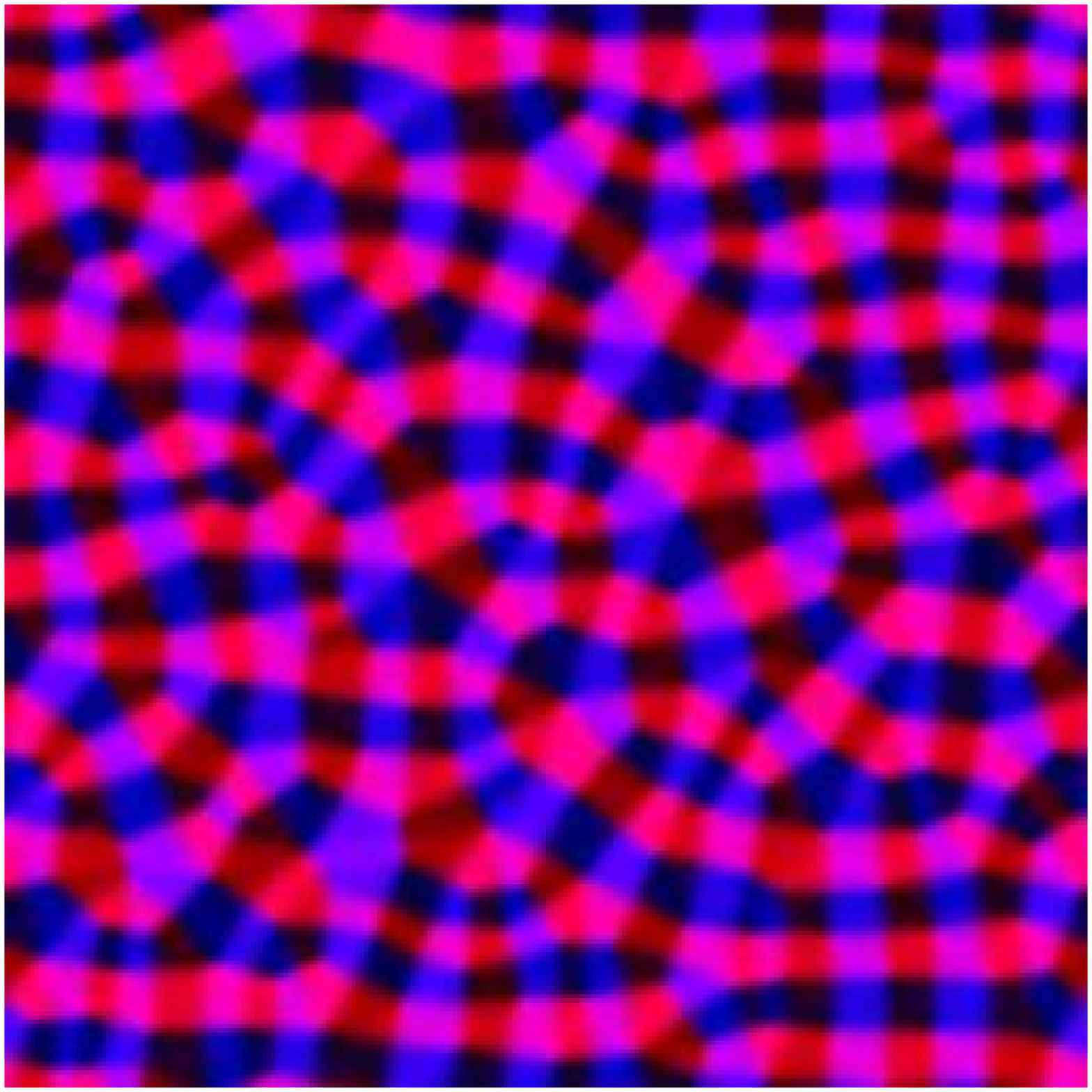}
\end{center}
%\vspace{-0.6cm}
\begin{center}
\includegraphics[width=5cm]{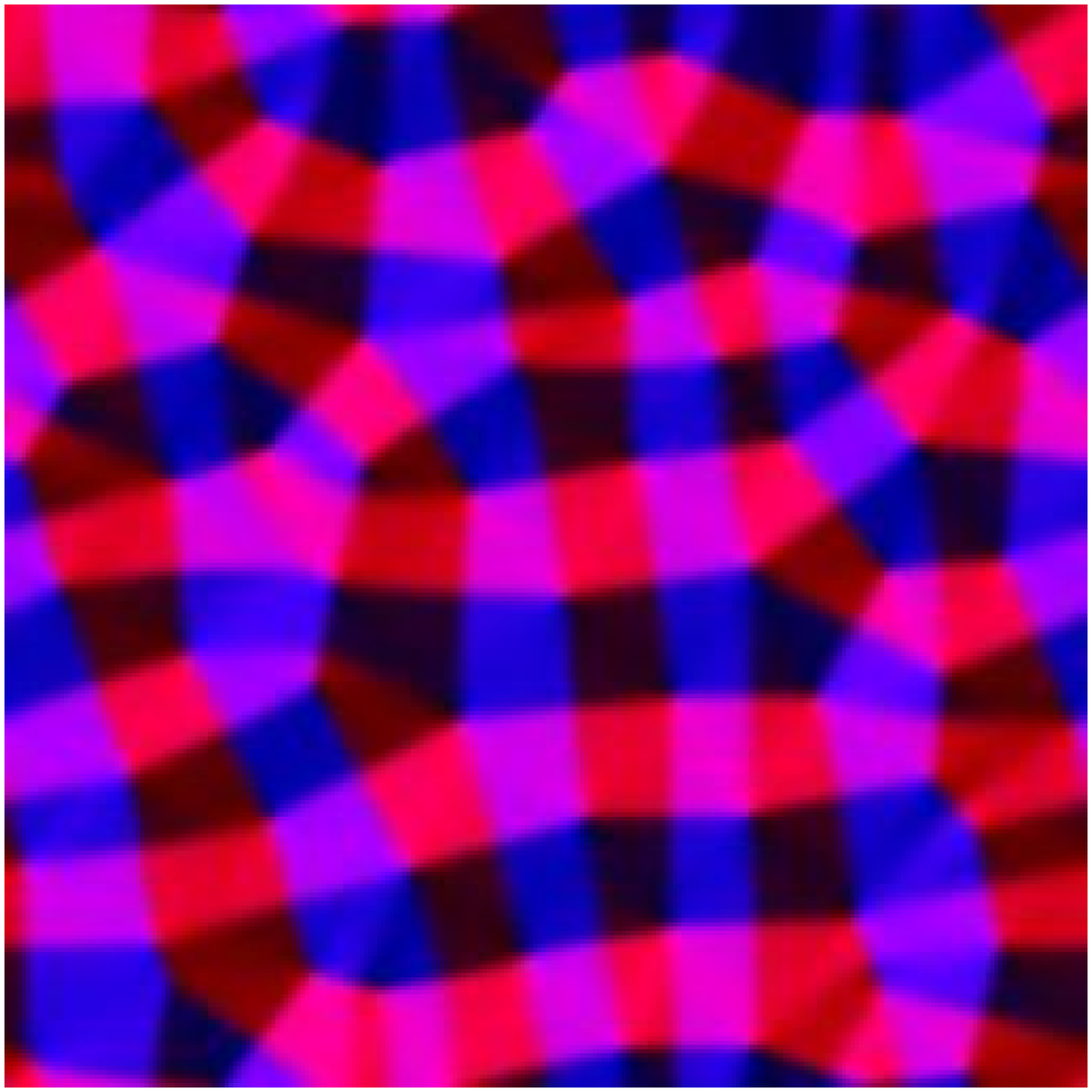}
\includegraphics[width=5cm]{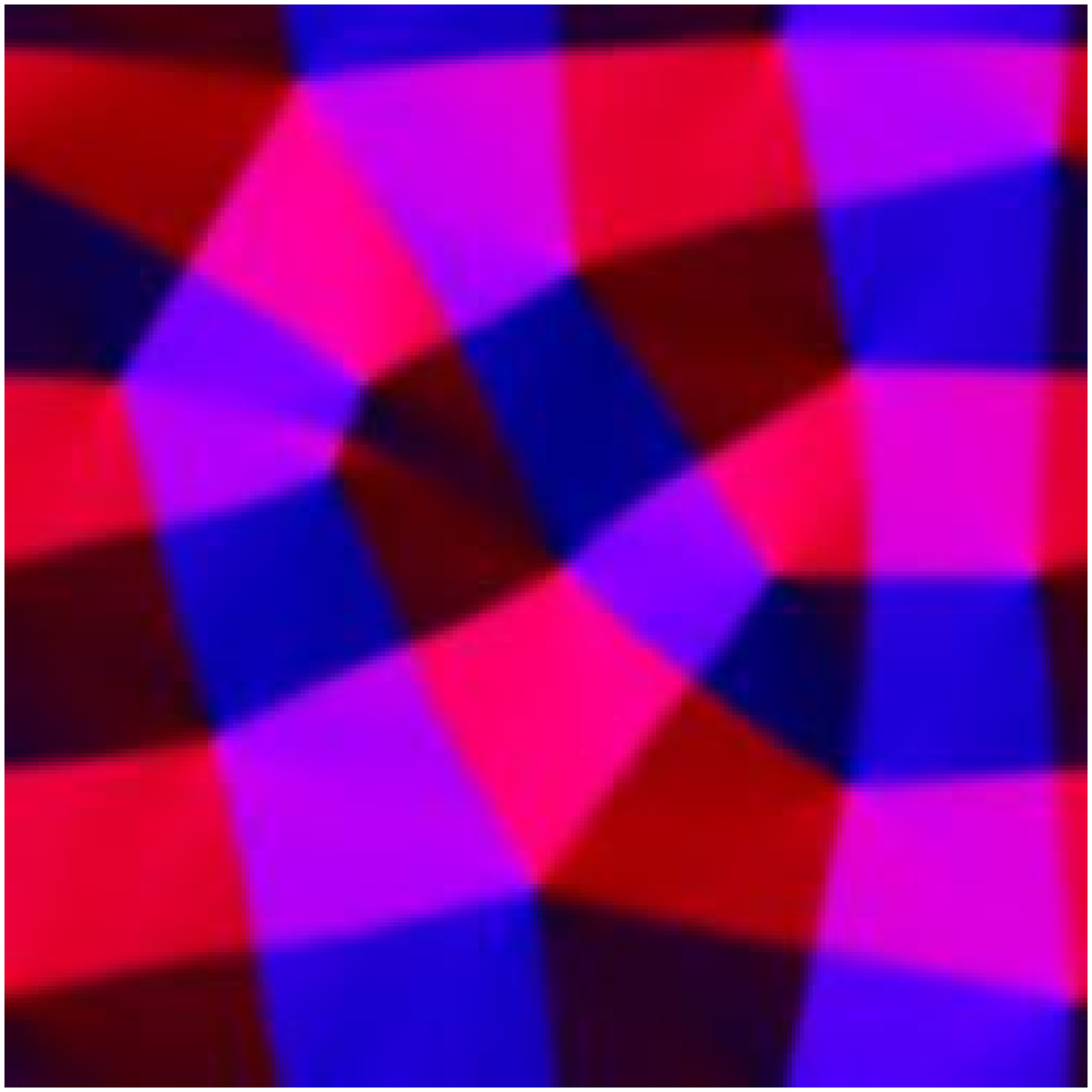}
\end{center}
\label{Fig:Coarse}
\caption{Coarsening of surface mounds, between subsequent panels time
  advances by a factor of 10. The local surface slope is represented
  by colours, facets appear as regions of uniform colour. Spontaneous
  facet formation, although no anisotropy is imposed, ${\bf J}_{\rm
  EST} ({\bf m}) = (1 - |{\bf m}|^2) \; {\bf m}$ (c.f.\ left panel in
  Fig.~\ref{Fig:ES}).} 
\end{figure}

\section{Examples of applications}
Continuum equations are now applied in two examples of very basic
choices in Eq.~(\ref{eq:surfdyn}). In some sense they highlight two
opposite regimes, (i) on very long length and time scales coarsening of
mounds in homoepitaxy, and (ii) on short scales resolving the lattice
structure in the transition from layer--by--layer to rough growth.

\subsection{Coarsening}
In some sense the simplest example of a surface dynamics equation is
the analogon to the classical XY--model in the sense of
Eq.~(\ref{eq:modelb})
\begin{equation}
\partial_t h = - (\nabla^2)^2 h - \nabla \cdot \left[ \left(1 - (\nabla
  h)^2 \right) \; \nabla h \right],
\label{eq:siegert}
\end{equation}
i.e., with the current function depicted in the left panel of
Fig.~\ref{Fig:ES}. An initially flat surface $h = const$ is linearly
unstable to fluctuations of wave lengths $\lambda > 2 \pi$ which
therefore initially grow exponentially. Once slopes of $O(1)$ are
reached the surface organises itself into a pattern of regular mounds
and troughs which keep the symmetry $h \leftrightarrow - h$ of
Eq.~(\ref{eq:siegert}). The in--plane isotropy of (\ref{eq:siegert})
is locally broken by the need to arrange the mounds and troughs into a
regular pattern and facets with sharp wedges form \cite{Politi00}.

Subsequently the pattern of mounds and troughs coarsens keeping a
single typical length scale and a statistically self similar pattern
at every time. The average mound size is found to increase with time
as $L \sim t^{1/3}$ \cite{MG00,S98}. In a weak sense this can be
proved as an upper bound for an appropriate time average of the length
scales $L$ observed up to a give time $t$ \cite{KY03}. Here only a
quick handwaving argument is given following \cite{RK97a}:

\begin{figure}[!ht]
\begin{center}
\includegraphics[width=8cm]{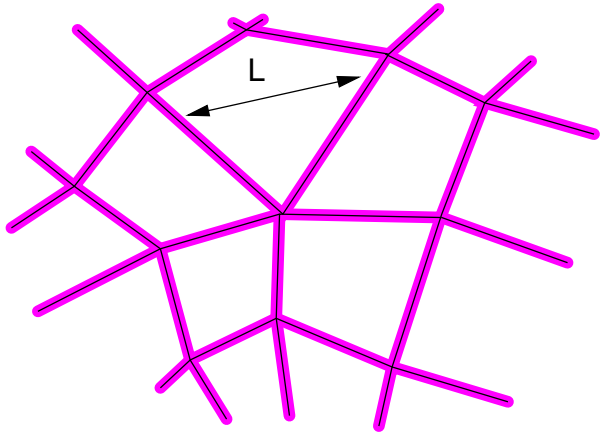}
\end{center}
\label{Fig:CSketch}
\caption{Sketch of facets and wedges, i.e., domains of constant slope
  and their separating boundaries, as obtained in the simulations
  shown in Fig.~\ref{Fig:Coarse}. The only macroscopic length scale is
  the domain size $L$.} 
\end{figure}

The spatial average of the interface width $W \equiv \langle h^2
\rangle$ (where $\langle h \rangle = 0$) increases due to
Eq.~(\ref{eq:siegert}) as
\begin{equation}
\frac{1}{2} \; \partial_t W^2 = \frac{1}{2} \; \partial_t \langle h^2
\rangle = - \langle (\Delta h)^2 \rangle + \langle (\nabla h)^2
\rangle - \langle |\nabla h|^4 \rangle.
\end{equation}
By the sketch of Fig.~\ref{Fig:CSketch} the contributions of these
terms are estimated to be of order $1/L$: The facets are flat and the
curvature is restricted to narrow strips of with $O(1)$ at the domain
boundaries, which have a relative weight of $1/L$. Also, here the
slopes are smaller than the stable value $|\nabla h| = 1$ attained at
the facets. So all terms give a contribution of $O(1/L)$ and because
of the stable slope $W \sim L$ the increase in width leads to the
estimate
\begin{equation}
\partial_t L^2 \sim \partial_t W^2 \sim 1/L, \; \; \; \mbox{integrated
  to} \; \; \; L \sim t^{1/3}.
\end{equation}
The same rough heuristic argument holds for anisotropic surface
currents with discrete minima, i.e., a discrete set of preferential
surface orientations, but the four--fold symmetry (right panel of
Fig.~\ref{Fig:ES}) plays a special role because two kinds of edges
occur, such that in general the assumption of a single length scale
$L$ in the system is wrong \cite{S98}.

\subsection{Roughening}
Layer by layer growth is technologically interesting because one can
monitor the subsequent deposition of layers by oscillations in the
surface width. It is nearly flat around integer fillings and the
roughness takes maxima at half integer fillings. During growth the
interface roughens and the width oscillations decay \cite{KKBW97}.
The Ehrlich--Schwoebel barrier and mounding instability is one
prominent mechanism destroying layer--by--layer growth, and so can be
any inhomogeneity in the deposition intensity $F_{\{ h \}}({\bf
  x},t)$. Although weaker, even the fluctuations of shot noise have
this effect, and it can be captured by a simple renormalisation group
calculation.

The simplest way to implement layer lattice effects into an continuum
surface equation like Eq.~(\ref{eq:surfdyn}) is pursued by the
conserved {\em sine--Gordon}--equation
\begin{equation}
\partial_t h({\bf x},t) = - K \Delta^2 h({\bf x},t) - V \Delta \sin
  \frac{2\pi h({\bf x},t)}{a_\perp} + F + \xi({\bf x},t). 
\label{eq:csg}
\end{equation}
% - \lambda \Delta (\nabla h)^2
Among other the surface current is driven to regions of incomplete
fillings such that $h({\bf x},t)/a_\perp$ is favoured to take integer
values \cite{RK97b}. The notation distinguishes between the lattice
constant normal ($a_\perp$) and parallel ($a_\|$) to the surface
terraces which will be used in the following analysis.

In a momentum shell renormalisation one can average over the
fluctuations caused by the short wavelength components of the noise,
$\xi = \bar \xi + \delta \xi$, where $\bar \xi$ only contains modes
of wavenumber smaller than some cutoff $\Lambda$ and $\delta \xi$ the
others. One now tries to approximate the dynamic equation for the
averaged height field $\bar h = \langle h \rangle_{\delta \xi}$ by an
equation like (\ref{eq:csg}) with suitably adjusted parameters. After
rescaling, an effective {\em renormalised} equation is obtained. Three
important physical interpretations can be obtained from the
renormalised parameters of Eq.~(\ref{eq:csg}).

\medskip
\noindent
(i) by the interplay of the diving force $F$ and the lattice potential
the up--down symmetry ($h \leftrightarrow -h$) is broken and a term $-
\lambda \Delta (\nabla h)^2$ is created.

\medskip
\noindent
(ii) The strength of the lattice potential $V$ is damped out decays
exponentially with coverage $\theta = Ft$ like $\propto \exp \left( -
a_\|^2 \! / \! \ell_{\rm D}^2  \sqrt{\theta} \right)$

\medskip
\noindent
(iii) In a finite size sample smaller than the so called {\em layer
  coherence} length $\tilde \ell = \ell_{\rm D}^2 / _\|$
renormalisation has to stop before the lattice potential has
decayed. Shot noise fluctuations cannot accumulate to the point where
width oscillations are completely blurred.

%\section{Disadvantages and open questions}
\section{Discussion}
In presenting the model derivations and examples of their application
it becomes clear that continuum models are a good tool for questions
where the inherent {\em simplifications} are justified. The insight
gained from partial differential equations is different from results
of more microscopic model. Last not least continuum equations can
handle larger systems and longer times than Monte Carlo simulations or
even Molecular Dynamics.

%\subsection{Lattice effects and discreteness}
In many cases the phenomena of particular technological interest cannot
be captured by continuum equations of the type described in this
lecture. Magic island shapes, surface reconstruction, step bunching,
formation of quantum dots and wires are just a few examples.

Even if one isn't interested in these inherently discrete phenomena
themselves they nevertheless carry over to the behaviour on larger
scales and a naive straightforward derivation of large scale continuum
models may be wrong. The implementation of noise for some questions,
such as roughening, is one such example. Also, the effective ``large
scale'' Ehrlich--Schwoebel barrier of a step is influenced by kinks
along the step which may serve as channels for downward incorporation
of a terrace adatom. Obtaining the correct effective barrier,
dependent on orientation and curvature of the step certainly is a
nontrivial task.

An important part of the surface current ${\bf J}$ flows along steps,
as has become clear in the above derivation. In ref.~\cite{PK00} the
step current due to kink ES barriers is derived as a function of the
surface orientation or slope. An open question is the role of step
curvature and therefore of surface curvature on the step current.

Also on the surface itself, the regularising term as it is derived in
Eq.~(\ref{eq:mullins}), should be derived from true nonequilibrium 
arguments. To do so, one would have to incorporate island nucleations,
the creation and annihilation of steps. One possible approach is to
introduce fields for the densities of adatoms, advacancies, steps,
kinks, and for their orientations. Locally they could be treated as
mean field densities with certain effective ``reaction'' rates, but in
their spatial interaction, by diffusion and advective transport, one
could obtain a conceptually more fundamental continuum model for
surface growth.

\section{Conclusion}
In this lecture an overview of continuum models for MBE surface growth 
has been given which includes
\begin{itemize}
\item the conceptual foundation of continuum models as a limit of more
  detailed small scale approaches,
\item heuristic derivations due to lack of rigorous or even exact ways
  to obtain them,
\item the structure and the symmetry of the equations obtained in that
  way,
\item two examples emphasising two opposite time limits in the growth
  process with the application of two contrary methods,
\item criticism of weak points in continuum modelling, open questions
  and some speculations about answers to them.
\end{itemize}
The main purpose was to highlight the physical background and give
some intuition of possible approaches to it as a basis for
mathematical treatments, both rigorous and numerical. It should be
seen as a complement of the lectures \cite{BiehlOW} and \cite{KrugOW}
which give a similar introduction to lattice and
Burton--Cabrera--Frank models respectively. Although continuum models
seem to fail in some important physical aspects they nevertheless are
a good tool and subject for mathematical approaches.

% ------------------------------------------------------------------------

\subsection*{Acknowledgement}
This work, as well as the participation at the workshop was supported
by SFB 611 (Singul\"are Ph\"anomene und Skalierung in mathematischen
Modellen) of Deutsche Forschungsgemeinschaft.
% ------------------------------------------------------------------------
\end{document}